\documentclass[prl,twocolumn,preprintnumbers,amsmath,amssymb]{revtex4-1}

\usepackage{amsmath}
\usepackage{graphicx}
\usepackage{dcolumn}
\usepackage{bm}

\DeclareMathOperator{\sign}{sign}

\begin{document}
 
\title{Attraction Controls the Inversion of Order by Disorder} 

\author{Fabio Leoni}
\author{Yair Shokef\footnote{\texttt{shokef@tau.ac.il}}}
\affiliation{School of Mechanical Engineering and Sackler Center for
  Computational Molecular and Materials Science, Tel-Aviv University, 
  Tel-Aviv 69978, Israel} 

\begin{abstract}
We show how including attraction in interparticle interactions
reverses the effect of fluctuations in ordering of a prototypical
artificial frustrated system. Buckled colloidal monolayers exhibit the
same ground state as the Ising antiferromagnet on a deformable
triangular lattice, but it is unclear if ordering in the two systems
is driven by the same geometric mechanism. By a real-space expansion
we find that for buckled colloids bent stripes constitute the stable
phase, whereas in the Ising antiferromagnet straight stripes are
favored. For generic pair potentials we show that attraction governs
this selection mechanism, in a manner that is linked to local packing
considerations. This supports the geometric origin of entropy in
jammed sphere packings and provides a tool for designing
self-assembled colloidal structures. 
\end{abstract}
  
\maketitle

Geometrically-frustrated systems cannot satisfy all local constraints,
and thus they remain disordered down to zero temperature
\cite{tarjus2005}.    
Subtle effects such as boundary conditions, lattice distortions, and
higher-order or long-range interactions can remove the degeneracy and
lead to an ordered ground-state.  
Alternatively, if entropy of fluctuations about each ground-state
configuration slightly varies, then the configuration with the highest
entropy \footnote{Formally, we should seek the configuration that
minimizes free energy rather than for the one that maximizes
entropy. However we show that in our case this is equivalent
\cite{supplmat}} will be thermodynamically selected in an effect
termed order by disorder
\cite{collins1997,wang2008,starykh2010,guruciaga2016}.   

\begin{figure}[t!]
\includegraphics[clip=true,width=8.5cm,height=5.24cm]{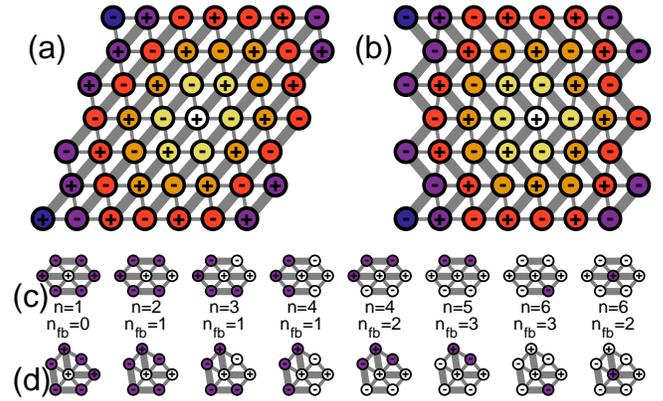}
\caption{(a) Straight and (b) bent configurations. 
For the Ising model we consider shells of neighbors around the central 
white particle, which are denoted with increasingly darker colors. 
Thicker lines correspond to frustrated bonds connecting particles at
the same state. 
For straight (c) and bent (d) stripes in the buckled monolayer we
consider different sets of $n$ free particles (white) with $n_{fb}$
frustrated bonds, while other particles (purple) are fixed.} 
\label{fig:stripes_shell}  
\end{figure}

Recently, frustration typical of antiferromagnetic spin
models has been studied in mesoscopic systems, composed of magnetic
islands \cite{wang2006,lammert2010,nisoli2013}, colloidal spheres
\cite{han2008,libal2006,ortiz2016,tierno2016} or elastic beams
\cite{kang2014,coulais2016}. The ability to visualize and manipulate
individual particles is also very useful to study glass formers
\cite{gokhale2016}, crystals, and gels \cite{lu2013}.   
For colloidal spheres confined between parallel walls, varying density
and plate separation (from one to two colloid diameters) a first-order
fluid freezing transition and discontinuous phase transitions between
layered, buckled and rhombic crystal structures occur
\cite{schmidt1996,schmidt1997}.
When the density approaches close packing, the monolayer buckles out
of its plane and neighboring spheres tend to touch opposite walls,
giving rise to effective antiferromagnetic interactions
\cite{shokef2009} and to glassy dynamics \cite{han}.
Multiple states with the same maximal density are obtained by
alternating straight stripes of up and down spheres
(Fig.~\ref{fig:stripes_shell}a) or by any set of zigzagging stripes
(Fig.~\ref{fig:stripes_shell}b).  

Slightly below this close-packing density it is difficult to find
analytic results regarding the thermodynamically stable phase.  
Instead, one can study an antiferromagnetic Ising model on a
deformable triangular lattice, which exhibits a highly-degenerate
ground state of randomly-zigzagging stripes at $T=0$, corresponding to
the close-packed density (at $\rho=\rho_c$) in the colloidal
system. 
The degeneracy of the ground state is removed for $T>0$
through an order-by-disorder effect: the thermodynamically stable
phase is set by differences in entropy of fluctuations around the
different ground-state configurations.    
In this model, fluctuations around the ground state are harmonic, thus
the phonon spectrum may be calculated and from it one finds that
straight stripes are selected \cite{shokef2011}.     
While this approach allows to compute the entropy of the system, it is 
unclear what is the mechanism behind the entropy selection of the
ground state and what is the full connection between this model and
confined colloidal systems, and jammed packings of particles in the
broader sense.     

In this Letter we compute in real-space coordinates the entropies of
the competing stripe configurations both in the deformable Ising
antiferromagnet and in the confined colloidal system for colloids
modeled with different pair potentials.
To compute the entropy of colloids modeled as hard-spheres we develop
a geometric approach related to that employed to estimate the free
volume of fcc vs hcp structures \cite{rudd1968,koch2005,radin2005}.      
While one could expect that repulsion would flatten out things and
give a preference for straight stripe, we find that generic
repulsive potentials give preference for bent stripes. 
In the Ising antiferromagnet straight stripes are favored, since
there attraction is included.   
This implies that attraction is responsible for flipping the sign 
of the order-by-disorder effect. 
 
\noindent{\it Ising Model}:
Each spin is linked to six neighbors by harmonic springs to form a
deformable triangular lattice, with the Hamiltonian given by the sum
over all nearest-neighbor pairs 
$\mathcal{H}=\sum_{\langle ij\rangle}\left[(1-\epsilon\delta
  r_{ij})J\sigma_i\sigma_j+\dfrac{K}{2}\delta r_{ij}^2\right].$
$\sigma_i=\pm 1$ and $\delta
r_{ij}=|\overrightarrow{r_i}-\overrightarrow{r_j}|-a$ are the spin and
relative position variables, respectively, $J>0$ is the
antiferromagnetic interaction strength, $a$ is the relaxed spring
length, $\epsilon>0$ is the rate at which the antiferromagnetic
interaction decreases linearly with distance, and $K$ the spring
stiffness. 
In the ground state, each plaquette deforms to an isosceles triangle 
with two shorter satisfied bonds and one longer frustrated
bond. Minimizing energy with respect to the head angle $\alpha$ of
these isosceles triangles relates $\alpha$ to the dimensionless ratio
$b=J\epsilon/Ka>0$ of the magnetoelastic interaction to the lattice
rigidity \cite{shokef2011}.  
At sufficiently low temperature (i.e. $k_BT\ll J$), spins cannot flip
and the Hamiltonian for straight and bent stripes configurations can
be expanded around mechanical equilibrium:
$\mathcal{H}=K\sum_{m,n}A_{m,n}q_{m}q_{n}$.
$\{q\}=\{u,v\}$ represents small displacements about the equilibrium
position of every spin, namely, $m$ and $n$ run from $1$ to $2N$, where
$N$ is the number of spins in the lattice. Since $b$ is set by
$\alpha$, the dimensionless matrix $A$ depends only on the deformation
angle $\alpha$ and on the zigzagging stripe realization
$\left\{\sigma_i\right\}$. 
The canonical partition function (up to multiplicative constants)
reads 
\begin{equation}\label{equ:Z}
Z =\hspace{-0.1cm}\displaystyle\int{\exp(-\beta
  K\sum_{m,n}A_{m,n}q_{m}q_{n})}d\{q\}\hspace{-0.1cm}=\hspace{-0.1cm}\left(\dfrac{\pi}{\beta
  K}\right)^N\hspace{-0.3cm}\|A\|^{-1/2} 
\end{equation}
where $\beta=1/T$, $\|A\|$ is the determinant of $A$ and we measure
temperature in units in which Boltzmann's constant is one. 
The entropy reads
\begin{equation}\label{equ:entropy}
S = -\dfrac{\beta}{Z}\left(\dfrac{\partial Z}{\partial\beta}\right)+\ln{Z}
 = N\left[1+\ln\left(\dfrac{\pi}{K\beta}\right)\right]-\dfrac{1}{2}\ln{(\|A\|)}.
\end{equation}

\begin{figure}[t!]
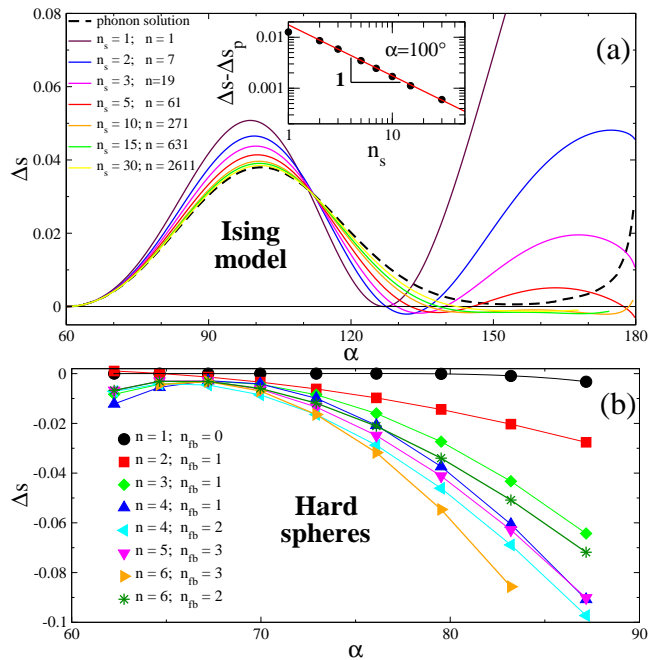

\begin{center}
\includegraphics[clip=true,width=8.5cm,height=4.73cm]{Fig2.eps}
\includegraphics[clip=true,width=8.5cm,height=3.91cm]{Fig3.eps}
\end{center}
\caption{Entropy difference per particle between straight and bent
  configurations vs deformation angle $\alpha$.  
(a) Ising antiferromagnet: Results with different numbers or free
  spins converge to the phonon solution $\Delta s_p$. Inset: distance
  between $\Delta s$ and $\Delta s_p$ at $\alpha=100^o$ vs $n_s$.  
(b) Hard-sphere monolayer for configurations in
  Figs.\ref{fig:stripes_shell}c,d. The line for $n=1$ is an analytic
  expression, lines for $n>1$ are guides to the eye.}      
\label{fig:entropy} 
\end{figure}

Computing the entropy of the system considering some particles free to 
move and all others fixed in their ground-state positions enables
us to analyze the contribution to the entropy coming from a specific
subset of particles. This requires finding a recursive relation for
$A$, which can be extended to an increasing number of free
particles. To this end we consider shells of particles around some
central particle for straight (Fig.~\ref{fig:stripes_shell}a) and bent 
(Fig.~\ref{fig:stripes_shell}b) configurations, and include the
fluctuation of all $n=1+3n_s(n_s-1)$ particles up to shell $n_s$ for
increasing $n_s$.  
Fig.~\ref{fig:entropy}a shows that the entropy difference per
particle $\Delta s=(S_{straight}-S_{bent})/n$ between
straight and bent configurations as obtained from our shell expansion
method converges to the exact phonon solution
\cite{shokef2011}. 
\emph{Considering only one particle free to move already gives a
  qualitative picture of the entropy difference.}  

\noindent{\it Buckled hard spheres:} 
Colloidal particles with short-range repulsion are usually
theoretically approximated as hard spheres \cite{han2008,han2008b}.
For hard spheres, momentum variables are irrelevant and, except for an
additive constant, the entropy reads $S_{N}=\ln(V_{N})$ where $V_{N}$
is the $3N$-dimensional phase-space volume available to the centers of
the $N$ spheres.  

\begin{figure}[t!]
\begin{center}
\includegraphics[clip=true,width=8.5cm,height=3cm]{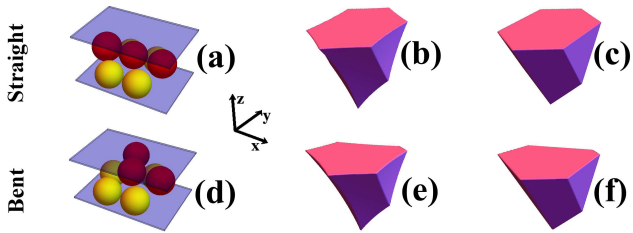}
\end{center}
\caption{Straight (a-c) and bent (d-f) configurations. (a,d) yellow
(red) spheres are in contact with the bottom (top) plane.  
Phase-space volume of one hard sphere free to move given by the
3-dimensional volume $V_1$, considering curved (b,e) and flat
(c,f) surfaces. Here $\alpha=70^o$ and $\delta/c=0.08$ to emphasize
the difference between curved and flat surfaces.}   
\label{fig:Volumes} 
\end{figure}

We compute the entropy of the straight and bent close-packed
configurations (see Fig.~\ref{fig:Volumes}) by directly calculating
their phase-space volume.      
To obtain the entropy of fluctuations about the close-packed state we
slightly decrease the density below close-packing by reducing the
radius $c$ of all spheres by $\delta\ll c$ \cite{supplmat}. 
In this way spheres have room to move, and following a free-sphere
expansion \cite{rudd1968}, we allow an increasing number $n$ of
contiguous spheres to move while the centers of the other $N-n$ remain
fixed in their close-packed positions. 

The shrinking of spheres, $c\rightarrow c-\delta$, implies the scaling
$V_{n}\sim\delta^{3n}$. 
$V_{n}$ is the phase space volume associated to the $n$ spheres free
to move, described by a $3n$-dimensional volume delimited by curved
surfaces.  
Considering one sphere free to move, the surface of the free volume
$V_{1}$ is described by rolling the free sphere in all possible ways
over its six neighbor spheres and over one confining wall
(Fig.~\ref{fig:Volumes}b,e).   
For $\delta\ll c$ it is possible to neglect the curvature of the
surfaces \cite{rudd1968,koch2005,radin2005}, thus obtaining linear
restrictions (planes) (Fig.~\ref{fig:Volumes}c,f), and then to compute 
$V_{1}$ as a 3-dimensional polyhedron, and for $n$ free spheres to
similarly compute $V_{n}$ as a 3n-dimensional polytope
\cite{supplmat,bueler2000}.  
$V_1$ is given by the Voronoi cell associated to the center of the
free sphere scaled by $2\delta$, and the entropy of straight and bent
configurations at this level coincide (as for fcc and hcp
\cite{radin2005}), but here only up to a deformation angle of
$\alpha^*=2\arctan\left(\sqrt{8/(7+\sqrt{33})}\right)\simeq 77^o$
\cite{supplmat}, after which bent-stripe entropy is higher than  
straight-stripe entropy, see Fig.~\ref{fig:entropy}b.
For bent stripes with $\alpha>\alpha^*$ when the free sphere is close
to the top wall the condition coming from one of the bottom spheres
becomes irrelevant~\cite{supplmat}. 
Therefore, \emph{in confinement already with one free sphere, there 
are local configurations of a collectively jammed state
\cite{torquato2010} with the same close-packed density which
differ in their local stability for slightly decreasing density.}  
We will show that this is valid also for larger $n$.

For $n=2$, the free volume can be computed through a $6$-dimensional
integral~\cite{radin2005}. 
Because the correlated motion of the free spheres, to find the free
volume for $n>2$ requires more sophisticated tools. 
For that we use a modified version of the Lasserre method
\cite{lasserre1983a,lasserre1983b} 
implemented in the VINCI code~\cite{vinci}.
We calculated the phase-space volumes $V_{n}$ for straight and bent
configurations up to $n=6$ spheres that are free to move (see
Fig.~\ref{fig:stripes_shell}c,d) corresponding 
to $18$-dimensional polytopes~\cite{supplmat}.
Figure~\ref{fig:entropy}b shows the entropy difference per
particle for different sets of $n$ contiguous spheres free to move
with $\delta/c=0.001$ (considering smaller values of $\delta$ did not
change the result) with the same number of frustrated bonds $n_{fb}$
in the straight and bent configuration. 
{\it We find that bent stripes are thermodynamically stable, which is
  the opposite from the Ising antiferromagnet result shown in
  Fig.~\ref{fig:entropy}a.} 
Increasing the number of free spheres, the entropy difference
increases, especially for large angles.  

From existing experiments on colloidal monolayers
\cite{han2008,yunker2014,han} it is hard to conclude if the ground
state has a preference for straight or bent stripes because the system
has to be annealed very slowly~\cite{shokef2009,shokef2011}.
From the experimental interparticle potential of NIPA colloids
\cite{han2008b} it is possible to see that a simple correction to the
hard-sphere potential can be described by a decreasing
exponential. However, this potential is in the same 
``quasi-universality'' class with hard spheres \cite{bacher2014},
hence no qualitative change in the results is expected.   
Softer interactions such as the hard-core soft-shoulder potential
does not change the preference for bent over straight stripes
\cite{supplmat}.     

\begin{figure}[t!]
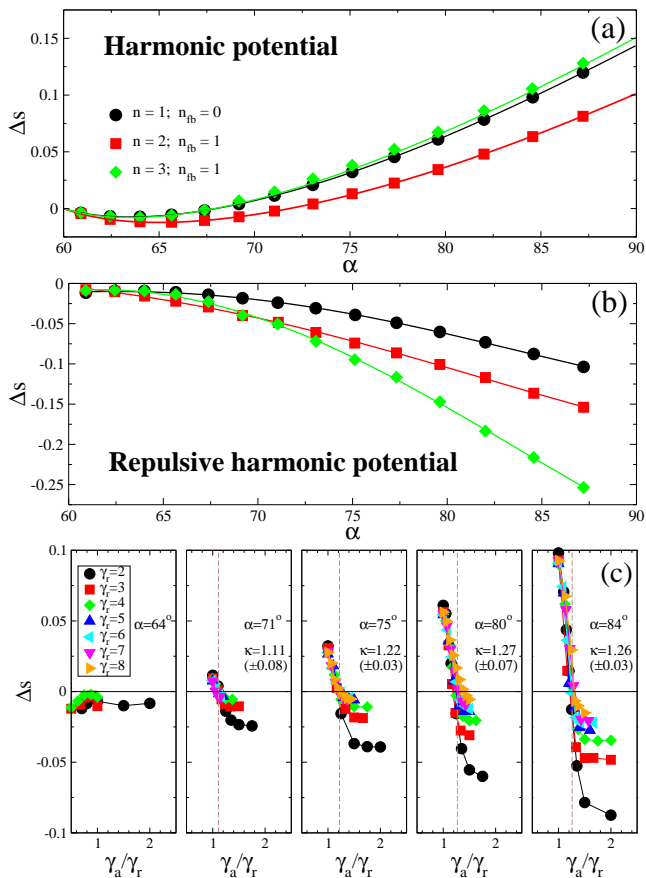

\begin{center}
\includegraphics[clip=true,width=8.5cm,height=3.53cm]{Fig5.eps}
\includegraphics[clip=true,width=8.5cm,height=3.53cm]{Fig6.eps}
\includegraphics[clip=true,width=8.3cm,height=4.45cm]{Fig7.eps}
\end{center}
\caption{Entropy difference per particle between straight and bent
  configurations vs deformation angle $\alpha$. 
  (a) Harmonic potential $U(\zeta)$.
  (b) Repulsive harmonic potential $U_r(\zeta)$.
  (c) Asymmetric potential $U_{asy}(\zeta)$: $\Delta s$ vs the ratio
  $\gamma_a/\gamma_r$ for different values of $\gamma_r$. 
  The curves are computed for $n=1$, $U_a^0=U_r^0$, and the vertical
  lines indicate $\kappa(\alpha)$.}    
\label{fig:entropy_U} 
\end{figure}

\noindent{\it Attraction}: 
Elasticity in the Ising model includes both attraction and repulsion
which contribute equally, while hard- or soft-sphere interactions used
to model colloids are purely repulsive.  
Adding attraction to repulsive colloids can induce a clustering phase
\cite{lu2006}, a solid-solid transition \cite{laar2016}, a glass-glass
transition \cite{voigtmann2011} and many other phenomena
\cite{gokhale2016}.  
To investigate the role of attraction in the entropy selection of the
ground state, we first consider a system of particles in the same
straight and bent positions as the buckled colloidal system with
particles interacting either through a harmonic potential
$U(\zeta)=U_0\zeta^2$ or through a purely repulsive harmonic potential
$U_r(\zeta)=U_0\zeta^2\cdot\theta(\zeta)$ where $\theta(\zeta)$ is the
Heaviside step function, $\zeta=(dr_0-dr)/(2\delta)$ and $U_0$ sets
the energy scale.
$dr_0=2c$ and $dr^2=dr_0^2+du^2+dv^2+dw^2$ with $\{u,v,w\}$ the
displacement around the equilibrium position. 
The entropy of straight and bent configurations for particles
interacting through $U(\zeta)$ can be exactly calculated
\cite{supplmat}, while for $U_r(\zeta)$ we use the canonical
partition function obtained by numerical integration \cite{supplmat}.  
Fig.~\ref{fig:entropy_U}a,b shows that the harmonic potential
$U(\zeta)$ gives a result qualitatively similar to that of the Ising
antiferromagnet (Fig.~\ref{fig:entropy}a) with a preference for
straight stripes.  
On the other hand, \emph{considering only the repulsive part of the
harmonic potential, using $U_r(\zeta)$, changes the preference to bent
stripes, as we found for hard spheres} (Fig.~\ref{fig:entropy}b). 

Considering a generalized repulsive potential
$U_r^{\gamma}(\zeta)=U_r^0\zeta^\gamma\cdot\theta(\zeta)$, which
reduces to the repulsive harmonic potential for $\gamma=2$ and gives
hard spheres of diameter $2(c-\delta)$ for
$\gamma\rightarrow\infty$ \footnote{$U_r^{\gamma}$ differs by a
  numerical prefactor from the commonly-used tunable soft repulsive
  potential \cite{liu2010,hecke2010,morse2014}, with
  $dr_0=\sigma_{ij}$ and $dr=r_{ij}$, and the hard-sphere limit
  obtained for $\alpha\rightarrow 0$ \cite{liu2010}.}, for
increasing $\gamma$, $\Delta s$ for one particle free to
move slowly approaches the hard-sphere result and always exhibits a
preference for bent stripes, implying that attraction is responsible
for the inversion of $\Delta s$ \cite{supplmat}.   
More generally we consider an asymmetric power-law potential
$U_{asy}(\zeta)=U_r^0\zeta^{\gamma_r}\theta(\zeta)+U_a^0(-\zeta)^{\gamma_a}\theta(-\zeta)$ 
through which we can tune both repulsion and attraction.  
We find a transition from bent to straight stripes by reducing the
attraction, that is for $\gamma_a/\gamma_r=\kappa(\alpha)$ (with
$0<1/\kappa<1$)~\cite{supplmat} (see Fig.~\ref{fig:entropy_U}c).
Note that for small angles, $\Delta s<0$ also for the symmetric
case. 
This is closely related to the geometric origin of the
preference for bent over straight stripes for $\alpha>\alpha^*$ for
one free hard sphere; considering a symmetric potential, like the
harmonic potential, means to always include the contribution
from all neighbors for all deformation angles, while considering a
purely repulsive potential (or properly reducing the attraction)
allows to disregard (or reduce enough) the contribution from one of
the bottom spheres for certain heights due to the same geometric
reasons as for hard spheres \cite{supplmat}.  
We conjecture that the same mechanism acts for more than one particle
free to move.        
It would be interesting to experimentally test this effect of
attraction and repulsion on the ground state, for colloidal systems
with frustration originating not only from confinement as we
considered above, but also possibly due to gravity \cite{ortiz2016},
optical trapping \cite{libal2006} or magnetic lattices
\cite{tierno2016}.  

\noindent{\it Conclusions}: We computed via real-space coordinates the
entropies for small fluctuations of the competing stripe
configurations both in the deformable Ising antiferromagnet and in the
colloidal monolayer for colloids modeled as hard or soft spheres.  
In the Ising antiferromagnet straight stripes are favored, while for
buckled colloids bent stripes are selected.
In many compact systems such as fcc and hcp using a harmonic potential
or a purely repulsive one doesn't change the result (for example by
lattice dynamical theory \cite{travesset2014,travesset}), yet we found
that it is fundamental.  
We found that attraction influences the ground-state selection
mechanism changing the sign of $\Delta s$, and we related it to local
geometric properties.     

Local geometry plays an important role in jamming
\cite{ashwin2009,morse2014}, even though it cannot give a complete
picture \cite{torquato2010}.    
Our results could provide insight into why some characteristics
of the jamming transition are related to local geometric properties
such as the mean number of nearest neighbors of Voronoi volumes and
mean number of constraints \cite{morse2014}.
Indeed, inverting $\Delta s$ for one free hard sphere corresponds to 
changing the number of nearest neighbors or the number of constraints.  
This mechanism is possible for dimensions $d\geq 3$ and it could be
related to the upper critical dimension for jamming suggested to be
equal to 3 \cite{morse2014}: shrinking spheres in a jammed state by
$\delta\ll c$, which is still a jammed state \cite{torquato2010}, we
can conjecture that there are many different local configurations with
the same density, the entropies of which after shrinking may differ
for each one of them. It would be interesting to test the relevance of
attraction in other buckled patterns, as for glassy states with
numerous coordination polyhedrons \cite{ma2015}.    

Our results provide a useful tool for designing self-assembled
colloidal systems: in a system with multiple possible configurations
differing in entropy, tuning the attractive and repulsive components
of the inter-particle potential can change the nature of the stable
phase.  
For example, DNA-coated colloids can be designed by controlling the
nucleotide sequences, coating densities and the attractive and
repulsive component \cite{A-U2012}, and attraction is responsible for
self-assembly of nanocrystal superlattices \cite{ye2013}. 

Predicting spontaneously-formed structures from properties of building 
blocks is another example of the role played by local geometry
\cite{damasceno2012}.  
The square lattice with quadratic interactions between
next-nearest-neighbour sites can be turned from stable to a highly
degenerate zigzag state by tuning the quadratic coefficient from
positive to negative \cite{mao2015}.
Crucial differences between random spring networks and jammed packings
caused by redundant constraints \cite{ellenbroek2015} could originate
from attraction. It could be interesting to study a possible
transition in a network of asymmetric-interacting points from the 
random spring to the jammed-packing behavior.  
Clearly, it would be interesting to test our theoretical predictions in
simulations and experiments and try to improve algorithms to find
random packings of jammed frictionless hard spheres.

We thank C. Calero, R. Golkov, Y. Han, E. O\u{g}uz, N. Segall,
A. Souslov, G. Tarjus and E. Teomy for helpful discussions.
This research was supported by the Israel Science
Foundation Grant No. 968/16.

\bibliography{paper_stripes.bib}
   
\clearpage
\onecolumngrid

\begin{center}
{\bf\large{Supplementary Material for ``Attraction Controls the
Inversion of Order by Disorder in Buckled Colloidal Monolayers''}}\\ 
\vspace{0.3cm}
Fabio Leoni and Yair Shokef\\
{\it School of Mechanical Engineering and Sackler Center for
  Computational Molecular and Materials Science, Tel-Aviv University, 
  Tel-Aviv 69978, Israel}
\end{center}

\setcounter{equation}{0}
\setcounter{figure}{0}
\setcounter{table}{0}
\setcounter{section}{0}
\makeatletter
\renewcommand{\theequation}{S\arabic{equation}}
\renewcommand{\thefigure}{S\arabic{figure}}
\renewcommand{\thetable}{S\arabic{table}}
\renewcommand{\thesection}{S\arabic{section}}

In the present supplementary material we provide detailed analytical
calculations behind the results presented in the main text of the
paper. 
In Sec. SI we compute the exact free volume available to the
center of a sphere in the straight and bent configurations near the
close-packing limit from which it is possible to calculate the entropy
of such systems. 
We give also some detail about our implementation of the VINCI code
used to compute the volume of high-dimensional polytopes.
In Sec. SII we compute the Hamiltonian and entropy of the soft
potential model in the small fluctuations approximation for straight 
and bent configurations.
In Sec. SIII we show that the sign of the entropy difference between
straight and bent stripes configurations of soft spheres interacting
through the square-shoulder repulsive potential is the same as for
hard spheres.
Finally, in Sec. SIV we define the asymmetric potential and show
the transition from straight to bent stripes as the contribution of
the attractive component of the potential is reduced compared to the
repulsive component.

\section{Volume calculation for one sphere free to move}

The centers of the spheres of one unit cell in the straight and bent
configurations (see Fig.~\ref{fig:Volumes}a, d) have coordinates
$\{(\pm c,\pm d,0)$, $(0,0,H)$, $(\pm 2c,0,H)\}$ and $\{(\pm c, -d,0)$,
$(c,d,0)$, $(2c,0,H)$, $(-c(3d^2-c^2)/(c^2+d^2),
d(3c^2-d^2)/(c^2+d^2), 0)$, $(0,0,H)$, $(-2c(d^2-c^2)/(c^2+d^2),
4c^2d/(c^2+d^2), H)\}$ respectively. The bottom and top confining
walls are at $z=-c$ and $z=H+c$ respectively. $c$ is the radius of
each sphere. Down and up sphere centers have z-coordinate 
$0$ and $H=\sqrt{3c^2-d^2}$, respectively.
The angle $\alpha$ ($60^o\leq\alpha\leq 90^o$) is the head angle of
the isosceles triangle obtained from the projection on the $xy$ plane
of the tilted equilateral triangle the corners of which are the
centers of the spheres with coordinates $(\pm c, d,0)$ and $(0,0,H)$
such that $c=d\tan{(\alpha/2)}$, which defines the parameter $d$.

\begin{figure}[t!]
\includegraphics[clip=true,width=12cm,height=5.6cm]{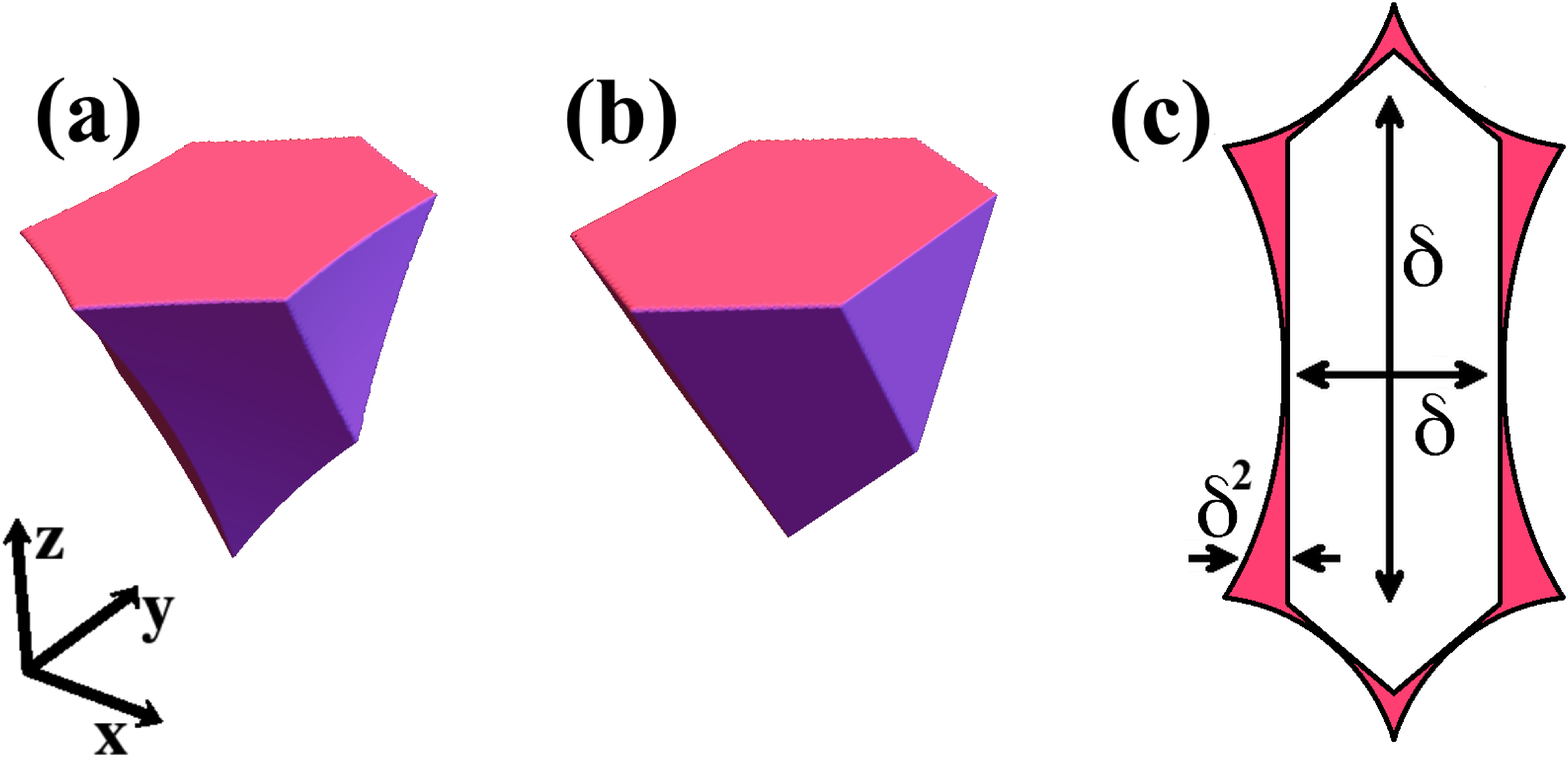}
\caption{Phase space volume of one hard sphere free to move in the
  straight stripe configuration given by the 3-dimensional volume
  $V_1$, considering curved (a) and flat (b) surfaces. Here
  $\alpha=70^o$ and $\delta/c=0.08$ to emphasize the difference
  between curved and flat surfaces. 
  (c) xy slice of the straight stripe free volume which shows the
  order in $\delta$ of the difference between curved and flat areas.}     
\label{fig:xy} 
\end{figure}

\begin{figure}[t!]
\includegraphics[clip=true,width=12cm,height=3.6cm]{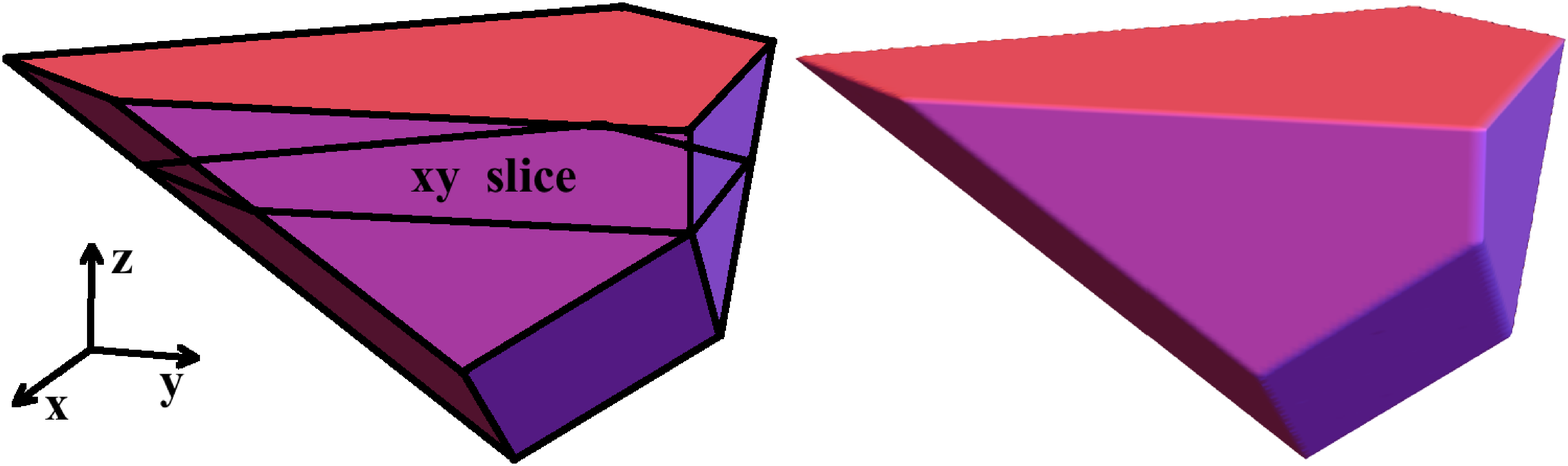}
\caption{Phase space volumes of one hard sphere free to move in the
  bent stripe configuration given by the 3-dimensional volume $V_1$,
  considering flat surfaces. Here 
  $\delta/c=0.08$, and $\alpha=90^o$, for which the
  difference between straight and bent stripe volumes is maximal.
  The left panel is a reproduction of the right one, but in which a
  $xy$ slice which when integrated along the $z$ axis gives $V_1$ is
  indicated.} 
\label{fig:xy_slice} 
\end{figure}

To compute the 3n-dimensional phase-space volume $V_n$ available to
the centers of $n$ spheres free to move, we slightly decrease the
density of straight and bent stripes configurations below the
close-packing value by reducing the radius $c$ of all spheres by
$\delta\ll c$.  
For angles close to $\alpha=60^o$, to avoid that free up (down)
spheres can be confined by the down (up) wall, $\delta$ should be
small enough. For one sphere free to move this condition is given by: 
$\delta/c<(1-\sqrt{1+cot^2(\alpha/2)}/2)$).    
The condition $\delta\ll c$ allows to neglect the curvature of the
surfaces which define the volume $V_n$ so that it is possible to
compute $V_n$ as a 3n-dimensional polytope. 
For $n=1$ this can be seen from simple geometric considerations; while
$V_{1}\sim\delta^3$, the difference between the volume $V_{1}$
computed considering curved surfaces and planar surfaces scales as
$\delta^4$ and thus may be neglected.   
This follows considering that the volume $V_{1}$ can be computed
integrating $xy$ slices along the $z$ direction
(see Figs.~\ref{fig:xy}, \ref{fig:xy_slice}).  
The perimeter of each slice of the curved polyhedron is composed of
arcs specified by the angle $\theta\sim\delta$ and the radius of
leading order $2c$ in $\delta$.
For each arc, the area of the corresponding segment (the area of a
sector minus the triangular piece) is given by
$2c^2(\theta-sin(\theta))\simeq c^2\theta^3/3\sim\delta^3$, so that
the difference between the area described by the curved and the flat
perimeter (indicated in pink in Fig.~\ref{fig:xy}c) is $\sim
\delta^3$. Considering the third dimension, $z$, gives the anticipated
result. 

\begin{figure}[t!]
\includegraphics[clip=true,width=11cm,height=7.56cm]{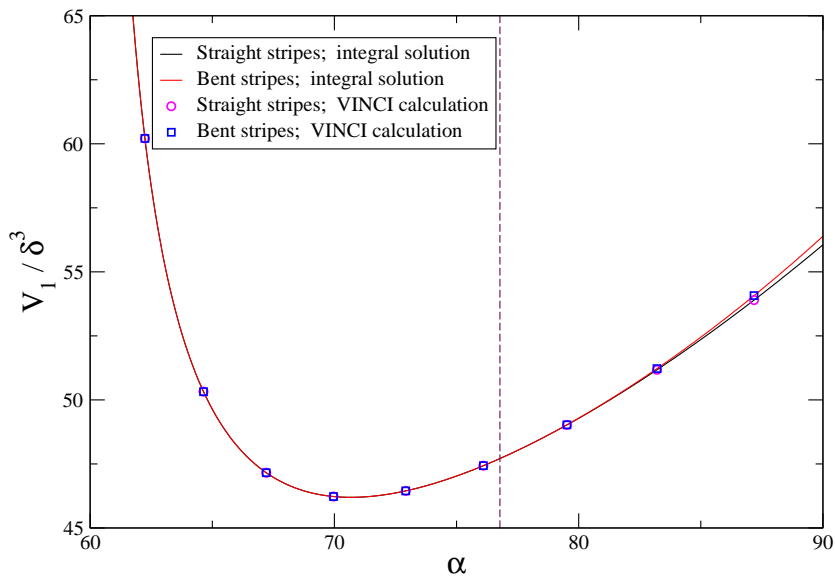}
\caption{The volume rescaled by $\delta^3$ available to the center of
  one sphere free to move in the straight
  ($V_1^{straight}/\delta^3$) and bent
  ($V_1^{bent}/\delta^3$) stripes configurations. The vertical
  dashed line corresponds to the threshold angle $\alpha^*\approx
  77^o$.}     
\label{fig:volumes_integrals} 
\end{figure}

The volume $V_1$ available to the center of one sphere free to
move is given by a $2\delta$-scaled Voronoi cell. 
The volume $V_1$ can be computed by integrating $xy$ slices
along the $z$ direction (Fig.~\ref{fig:xy_slice}).
For the straight stripes configuration, because the symmetry of $xy$
slices under the transformations $x\rightarrow -x$ and $y\rightarrow
-y$, we can just compute the area of the slice in one quadrant and
multiply it by 4 so that we obtain
\begin{equation}
\begin{array}{lll}
V_1^{straight} & = &
4\displaystyle\int_{H-\tfrac{4c}{H}\delta}^{H-\tfrac{2c}{H}\delta}dz\int_{\tfrac{H^2}{c}-\tfrac{Hz}{c}-4\delta}^{0}\left(\dfrac{c}{d}x-\dfrac{H^2-Hz-4c\delta}{d}\right)dx+4\int_{H-\tfrac{2c}{H}\delta}^{H+\delta}dz\int_{-2\delta}^{0}\left(\dfrac{c}{d}x-\dfrac{H^2-Hz-4c\delta}{d}\right)dx\\
&&\\
& = &
\dfrac{4c}{d}\left(6+\dfrac{H}{c}+\dfrac{28}{3}\dfrac{c}{H}\right)\delta^3=4
\left(6\tan\left(\dfrac{\alpha}{2}\right)+\sqrt{3\tan^2\left(\dfrac{\alpha}{2}\right)-1}+\dfrac{28\tan^2\left(\dfrac{\alpha}{2}\right)}{3\sqrt{3\tan^2\left(\dfrac{\alpha}{2}\right)-1}}\right)\delta^3
\end{array}
\end{equation}
From geometric considerations it is possible to see that in the bent
stripe configuration, for angles $\alpha$ larger than some threshold
value $\alpha^*$, the sphere centered in $(c,d,0)$ becomes irrelevant
in order to delimit the volume available to the free sphere for
$z>z^*$ with $z^*=H+2(d^2-c^2)\delta/(cH)$.  
Therefore, $\alpha^*$ corresponds to the extreme case in which the
previous inequality becomes an equality, that is for $z=H+\delta$ 
\begin{equation}
H+\delta=H+2(d^2-c^2)\delta/(cH)
\end{equation}
that can be written as
\begin{equation}
7\cot^2\left(\dfrac{\alpha^*}{2}\right)-4\cot^4\left(\dfrac{\alpha^*}{2}\right)-1=0
\end{equation}
hence
\begin{equation}
\alpha^*=2\arctan\left(\sqrt{\dfrac{8}{7+\sqrt{33}}}\right)\simeq 77^o
\end{equation}
The contributions to the volume $V_1^{bent}$ coming from angles
$\alpha<\alpha^*$ and $\alpha>\alpha^*$ have to be computed separately
because the $xy$ slices to be integrated along the $z$ axis are
described by polygons with a different number of edges (see
Fig.~\ref{fig:xy_slice}). So that we have
\begin{equation}\label{equ:V_s}
\begin{array}{lll}
V_1^{bent}(\alpha<\alpha^*) & = &
\displaystyle\int_{H-\tfrac{4c}{H}\delta}^{H-\tfrac{2c}{H}\delta}dz\Bigg[\int_{-\tfrac{2cH}{c^2+d^2}\left(z-H+\tfrac{4c}{H}\delta\right)}^{\tfrac{H(c^2-d^2)}{c(c^2+d^2)}\left(z-H+\tfrac{4c}{H}\delta\right)}\left(\dfrac{c(3d^2-c^2)}{dH^2}x+\dfrac{(c^2+d^2)}{dH}z+\dfrac{(c^2+d^2)(4c\delta-H^2)}{dH^2}\right)dx\\
&&\\
&& +\displaystyle\int_{\tfrac{H(c^2-d^2)}{c(c^2+d^2)}\left(z-H+\tfrac{4c}{H}\delta\right)}^{4\delta+\tfrac{H}{c}z-\tfrac{H^2}{c}}\dfrac{1}{d}(-x+Hz-H^2+4c\delta)dx\Bigg]+\displaystyle\int_{H-\tfrac{2c}{H}\delta}^{H+\delta}dz\Bigg[\int_{-\tfrac{2cH}{c^2+d^2}\left(z-H+\tfrac{4c}{H}\delta\right)}^{-\tfrac{2cH}{c^2+d^2}\left(z-H+\tfrac{c^2+d^2}{cH}\delta\right)}\\
&&\\
&& \left(\dfrac{c(3d^2-c^2)}{dH^2}x+\dfrac{(c^2+d^2)}{dH}z+\dfrac{(c^2+d^2)(4c\delta-H^2)}{dH^2}\right)dx+\displaystyle\int_{-\tfrac{2cH}{c^2+d^2}\left(z-H+\tfrac{c^2+d^2}{cH}\delta\right)}^{\tfrac{2cH}{c^2+d^2}\left(z-H+\tfrac{H}{c}\delta\right)}\\
&&\\
&& \dfrac{1}{2cd}((d^2-c^2)x+2(c^2+d^2)\delta)dx+\displaystyle\int_{\tfrac{2cH}{c^2+d^2}\left(z-H+\tfrac{H}{c}\delta\right)}^{2\delta}\dfrac{1}{d}(-cx+Hz-H^2+4c\delta)dx\Bigg]\\
&&\\
&&
-\displaystyle\int_{H-\tfrac{4c}{H}\delta}^{H+\delta}dz\int_{-\tfrac{2cH}{c^2+d^2}\left(z-H+\tfrac{4c}{H}\delta\right)}^{0}\dfrac{1}{d}(-cx-Hz+H^2-4c\delta)dx+\dfrac{1}{4}V_1^{straight}\\
&&\\
&& =\dfrac{4c}{d}\left(6+\dfrac{H}{c}+\dfrac{28}{3}\dfrac{c}{H}\right)\delta^3\equiv V_1^{straight}
\end{array}
\end{equation}
and
\begin{equation}\label{equ:V_b}
\begin{array}{lll}
V_1^{bent}(\alpha>\alpha^*) & = &
\displaystyle\int_{H-\tfrac{4c}{H}\delta}^{H-\tfrac{2c}{H}\delta}dz\Bigg[\int_{-\tfrac{2cH}{c^2+d^2}\left(z-H+\tfrac{4c}{H}\delta\right)}^{\tfrac{H(c^2-d^2)}{c(c^2+d^2)}\left(z-H+\tfrac{4c}{H}\delta\right)}\left(\dfrac{c(3d^2-c^2)}{dH^2}x+\dfrac{(c^2+d^2)}{dH}z+\dfrac{(c^2+d^2)(4c\delta-H^2)}{dH^2}\right)dx\\
&&\\
&& +\displaystyle\int_{\tfrac{H(c^2-d^2)}{c(c^2+d^2)}\left(z-H+\tfrac{4c}{H}\delta\right)}^{4\delta+\tfrac{H}{c}z-\tfrac{H^2}{c}}\dfrac{1}{d}(-cx+Hz-H^2+4c\delta)dx\Bigg]+\displaystyle\int_{H-\tfrac{2c}{H}\delta}^{H+\delta}dz\int_{-\tfrac{2cH}{c^2+d^2}\left(z-H+\tfrac{4c}{H}\delta\right)}^{-\tfrac{2cH}{c^2+d^2}\left(z-H+\tfrac{c^2+d^2}{cH}\delta\right)}\\
&&\\
&& \left(\dfrac{c(3d^2-c^2)}{dH^2}x+\dfrac{(c^2+d^2)}{dH}z+\dfrac{(c^2+d^2)(4c\delta-H^2)}{dH^2}\right)dx\\
&&\\
&& +\displaystyle\int_{H-\tfrac{2c}{H}\delta}^{H+\tfrac{2(d^2-c^2)}{aH}\delta}dz\Bigg[\displaystyle\int_{-\tfrac{2cH}{c^2+d^2}\left(z-H+\tfrac{c^2+d^2}{cH}\delta\right)}^{\tfrac{2cH}{c^2+d^2}\left(z-H+\tfrac{H}{c}\delta\right)}\dfrac{1}{2cd}((d^2-c^2)x+2(c^2+d^2)\delta)dx\\
&&\\
&& +\displaystyle\int_{\tfrac{2cH}{c^2+d^2}\left(z-H+\tfrac{H}{c}\delta\right)}^{2\delta}\dfrac{1}{d}(-cx+Hz-H^2+4c\delta)dx\Bigg]+\displaystyle\int_{H+\tfrac{2(d^2-c^2)}{cH}\delta}^{H+\delta}dz\displaystyle\int_{-\tfrac{2cH}{c^2+d^2}\left(z-H+\tfrac{c^2+d^2}{cH}\delta\right)}^{2\delta}\\
&&\\
&& \dfrac{1}{2cd}((d^2-c^2)x+2(c^2+d^2)\delta)dx-\displaystyle\int_{H-\tfrac{4c}{H}\delta}^{H+\delta}dz\int_{-\tfrac{2cH}{c^2+d^2}\left(z-H+\tfrac{4c}{H}\delta\right)}^{0}\dfrac{1}{d}(-cx-Hz+H^2-4c\delta)dx\\
&&\\
&& +\dfrac{1}{4}V_1^{straight}=\dfrac{\delta^3}{3c^2dH(c^2+d^2)}\Big[174c^6+88c^4d^2+18c^2d^4-8d^6+H(87c^5+47c^3d^2+12cd^4)\Big].
\end{array}
\end{equation}
In Fig.~\ref{fig:volumes_integrals} we show the volumes
$V_1^{straight}$ and $V_1^{bent}$ of one sphere free to
move for straight and bent stripes configurations, respectively,
rescaled by $\delta^3$ for several values of the deformation angles
$\alpha$.  

To compute the volume of high dimensional polytopes we use a modified
version of the Lasserre method \cite{lasserre1983a,lasserre1983b}
implemented in the VINCI code \cite{vinci}. 
Lasserre's method is a signed decomposition method which uses a half
space representation to describe each polytope. The modified version
implemented in \cite{vinci} incorporates a detection method of
simplicial faces and a storing/reusing scheme for face volumes which
can make a big use of computer memory (which increases exponentially
with the system size saturating the 384 GB of RAM of the computer we
used already for $n=5$, and forcing to use slow swap memory for $n=6$)
improving the efficiency of the original algorithm.   

In Fig.~\ref{fig:volumes_integrals} we compare the analytic expression
of volumes $V_1^{straight}$ and $V_1^{bent}$ we obtained
from direct integration in Eqs.~(\ref{equ:V_s},\ref{equ:V_b}) with
points we get from the VINCI code.   
From it we can see that the VINCI code is properly implemented in our
code and that for $\alpha>\alpha^*$ the curves corresponding to straight
and bent stripes split.

\vspace{0.3cm}

\begin{center}
{\bf SII. Soft potential model}
\end{center}

We consider the ground state configurations given by straight and bent
stripes in the soft potential model given by $U$ (harmonic potential),
$U_r$ (repulsive harmonic potential) and $U_r^{\gamma}$ (repulsive
power-law potential). The mechanical equilibrium position of particles
are given by $\{x_i,y_i,z_i\}$, and the displacement about these
positions are $\{u_i,v_i,w_i\}$.  
We consider small displacement around mechanical equilibria. The
distance between particles $i$ and $j$ is given by 
\begin{equation}\label{equ:dr2}
dr^2=(dx+du)^2+(dy+dv)^2+(dz+dw)^2=dr_0^2+2(dxdu+dydv+dzdw)+du^2+dv^2+dw^2
\end{equation}
where $dx=x_i-x_j$, $dy=y_i-y_j$, $dz=z_i-z_j$, $du=u_i-u_j$,
$dv=v_i-v_j$, $dw=w_i-w_j$, and $dr_0=(dx^2+dy^2+dz^2)^{1/2}=2c$ is the
equilibrium separation between the particles.
Ignoring the linear terms in $du$, $dv$ and $dw$ because we will
expand around mechanical equilibria (as in ref.~\cite{shokef2011}), we
write   
\begin{equation}\label{equ:dr2b}
dr^2=dr_0^2+du^2+dv^2+dw^2
\end{equation}
Because terms in the harmonic expansion of the Hamiltonian contain
also terms linear in $dr$, we take the square root of
Eq.~(\ref{equ:dr2}) and expand to harmonic order 
\begin{equation}\label{equ:dr}
dr=dr_0+\dfrac{du^2}{2dr_0}\left(1-\dfrac{dx^2}{dr_0^2}\right)+\dfrac{dv^2}{2dr_0}\left(1-\dfrac{dy^2}{dr_0^2}\right)+\dfrac{dw^2}{2dr_0}\left(1-\dfrac{dz^2}{dr_0^2}\right)-\dfrac{dxdydudv}{dr_0^3}-\dfrac{dxdzdudw}{dr_0^3}-\dfrac{dydzdvdw}{dr_0^3}
\end{equation}

The Hamiltonians $\mathcal{H}$, $\mathcal{H}_r$ and
$\mathcal{H}_r^{\gamma}$ associated to the potentials $U$, $U_r$ and
$U_r^{\gamma}$ respectively, are defined as 
\begin{equation}\label{equ:HamiltonS0}
\left\{\begin{array}{ccl}
\mathcal{H} & = &
U_0\sum_{m,n,p}\sum_{l} \zeta_l^2\\
\mathcal{H}_r & = &
U_0\sum_{m,n,p}\sum_{l} \zeta_l^2\cdot\theta(\zeta_l)\\
\mathcal{H}_r^{\gamma} & = &
U_0\sum_{m,n,p}\sum_{l}\zeta_l^\gamma\cdot\theta(\zeta_l)
\end{array}\right.
\end{equation}
where $\zeta_l=(dr_0-dr_l)/(2\delta)$. The usual sum over
nearest-neighbor pairs $\langle i,j,k\rangle$ is here replaced by all
particles $(m,n,p)$ and the label $l$ is associated to the additional
sum over the neighbors each particle has.
The range of values over which these sums are performed is specified
in the next sections for straight and bent configurations. There
we consider the Hamiltonian $\mathcal{H}$ for which the exact
result of the associated entropy can be written following the approach
specified for the Ising model in the main text in
Eqs.~(\ref{equ:Z},\ref{equ:entropy}). 
In particular, we compute the associated matrix $A$ for a specific set
of 1, 2 and 3 particles free to move according to its definition as:
$\mathcal{H}=U_0/(4\delta^2)\sum_{m,n}A_{m,n}q_mq_n$ where $K$ is
replaced by $U_0/(4\delta^2)$. We obtain the entropy of straight and
bent stripes configurations associated to the Hamiltonian
$\mathcal{H}_r$ and $\mathcal{H}_r^{\gamma}$ by directly integrating
the related partition function.


\subsection{Straight Stripes}

Performing the sum over the index $l=1,2,3,4,5,6$ associated to the
six neighbors of the central particle, we obtain (up to additive
constants) 
\begin{equation}\label{equ:HamiltonS}
\mathcal{H}=\dfrac{U_0}{4\delta^2}\sum_{m,n,p}\Big\{-4c(dr_1+dr_2+dr_3+dr_4+dr_5+dr_6)+(dr_1^2+dr_2^2+dr_3^2+dr_4^2+dr_5^2+dr_6^2)\Big\}+\mathcal{H}_{wall}
\end{equation}
where $\mathcal{H}_{wall}$ is the contribution to the Hamiltonian
coming from the interaction of the central particle with the wall.
The indices and positions at mechanical equilibrium of the central
particle and its six neighbors are given in Table S1 in which
$c=d\tan(\alpha/2)$ with $60^o<\alpha<90^o$ and  
$H=\sqrt{3c^2-d^2}$ with $0<H<\sqrt{2}c$.
\begin{table}[h!]
\label{tab:straight}
\begin{center}
\begin{tabular}{|c|*{4}{c|}}
\hline
\mbox{\it l} & index & x & y & z\\
\hline
0 & (m,n,p)       & 0  & 0 & H\\

1 & (m+1,n,p)     & 2c & 0 & H\\

2 & (m+1,n+1,p-1) & c  & d & 0\\

3 & (m-1,n+1,p-1) & -c & d & 0\\ 

4 & (m-1,n,p)   & -2c  & 0 & H\\

5 & (m-1,n-1,p-1) & -c & -d & 0\\

6 & (m+1,n-1,p-1) & c & -d & 0\\
\hline
\end{tabular}
\hspace{0.5cm}
\begin{tabular}{|c|*{3}{c|}}
\hline
particles & dx & dy & dz\\
\hline
1,0 & 2c & 0 & 0\\

2,0 & c  & d & -H\\

3,0 & -c & d & -H\\ 

4,0 & -2c  & 0 & 0\\

5,0 & -c & -d & -H\\

6,0 & c & -d & -H\\
\hline
\end{tabular}
\caption{Indices and positions of the central particle $0$ and his six
  neighbours $l=1,...,6$ (left) and distances between the neighbouring
  particles and the central one (right) in the unit cell of the straight
  stripe configuration.} 
\end{center}
\end{table}

Using Eqs.~(\ref{equ:dr2b}, \ref{equ:dr}) and the position of
particles in Table S1, noting that $du_l=u_l-u_0$, $dv_l=v_l-v_0$ and
$dw_l=w_l-w_0$, and absorbing the terms with indices $l=4,5,6$ in the
terms with indices $l=1,2,3$, we can rewrite Eq.~(\ref{equ:HamiltonS})
as  
\begin{equation}
\begin{array}{ll}
\mathcal{H}= &
\dfrac{U_0}{2\delta^2}\displaystyle\sum_{m,n,p}\Bigg\{\dfrac{3}{2}u_0^2-u_0u_1-\dfrac{1}{4}(u_0u_2+u_0u_3)-\dfrac{d}{4c}(u_0v_2+u_2v_0-u_0v_3-u_3v_0)\\
&\\
&+\dfrac{H}{4c}\sign(dz_0-H/2)(u_0w_2-u_0w_3+u_2w_0-u_3w_0)+\dfrac{d^2}{2c^2}v_0^2\\
&\\
&-\dfrac{d^2}{4c^2}(v_0v_2+v_0v_3)+\dfrac{Hd}{4c^2}\sign(dz_0-H/2)(v_0w_2+v_2w_0+v_0w_3+v_3w_0)\\
&\\
&+\left(2-\dfrac{d^2}{2c^2}\right)w_0^2-\dfrac{H^2}{4c^2}(w_0w_2+w_0w_3)\Bigg\}\\
\end{array}
\end{equation}
where $\sign(\cdot)$ is the usual sign function and it takes into
account the fact that the sign of the contributions to the Hamiltonian
coming from combination of variables $u_iw_j$ and $v_iw_j$ depends on
the particle $0$ being up or down. 
In the previous expression of the Hamiltonian, we considered the
contribution of the interaction with the wall given by
$\mathcal{H}_{wall}=U_0w_0^2dz^2_{wall}/(16c^2\delta^2)$ with $dz_{wall}=2c$. 
For bent configuration the contribution of $\mathcal{H}_{wall}$ is the
same as for the straight configuration.
In the following we specify the expression of the matrix $A$ for 1, 2
and 3 particles free to move, $A_s^{0}$, $A_s^{01}$ and $A_s^{012}$
respectively, where the subscript $s$ stays for straight and the
superscript indicates the particles as specified in Table S1, which are
free to move.
\begin{equation}
A_s^0=\left(\begin{array}{ccc}
3 & 0 & 0\\
&&\\
0 & \dfrac{d^2}{c^2} & 0\\
&&\\
0 & 0 & 4-\dfrac{d^2}{c^2}\\
\end{array}\right)
\end{equation}
\begin{equation}
A_s^{01}=\left(\begin{array}{cccccc}
3 & 0 & 0 & -1 & 0 & 0\\
&&&&&\\
0 & \dfrac{d^2}{c^2} & 0 & 0 & 0 & 0\\
&&&&&\\
0 & 0 & 4-\dfrac{d^2}{c^2} & 0 & 0 & 0\\
&&&&&\\
-1 & 0 & 0 & 3 & 0 & 0\\
&&&&&\\
0 & 0 & 0 & 0 & \dfrac{d^2}{c^2} & 0\\
&&&&&\\
0 & 0 & 0 & 0 & 0 & 4-\dfrac{d^2}{c^2}\\
\end{array}\right)
\end{equation}
\begin{equation}
A_s^{012}=\left(\begin{array}{ccccccccc}
3 & 0 & 0 & -1 & 0 & 0 & -\dfrac{1}{4} & -\dfrac{d}{4c} & \dfrac{H}{4c}\\
&&&&&&&&\\
0 & \dfrac{d^2}{c^2} & 0 & 0 & 0 & 0 & -\dfrac{d}{4c} &
-\dfrac{d^2}{4c^2} & \dfrac{dH}{4c^2}\\
&&&&&&&&\\
0 & 0 & 4-\dfrac{d^2}{c^2} & 0 & 0 & 0 & \dfrac{H}{4c} &
\dfrac{dH}{4c^2} & -\dfrac{H^2}{4c^2}\\
&&&&&&&&\\
-1 & 0 & 0 & 3 & 0 & 0 & -\dfrac{1}{4} & \dfrac{d}{4c} & -\dfrac{H}{4c}\\
&&&&&&&&\\
0 & 0 & 0 & 0 & \dfrac{d^2}{c^2} & 0 & \dfrac{d}{4c} &
-\dfrac{d^2}{4c^2} & \dfrac{dH}{4c^2}\\
&&&&&&&&\\
0 & 0 & 0 & 0 & 0 & 4-\dfrac{d^2}{c^2} & -\dfrac{H}{4c} &
\dfrac{dH}{4c^2} & -\dfrac{H^2}{4c^2}\\
&&&&&&&&\\
-\dfrac{1}{4} & -\dfrac{d}{4c} & \dfrac{H}{4c} & -\dfrac{1}{4} &
\dfrac{d}{4c} & -\dfrac{H}{4c} & 3 & 0 & 0\\
&&&&&&&&\\
-\dfrac{d}{4c} & -\dfrac{d^2}{4c^2} & \dfrac{dH}{4c^2} & \dfrac{d}{4c} &
-\dfrac{d^2}{4c^2} & \dfrac{dH}{4c^2} & 0 & \dfrac{d^2}{c^2} & 0\\
&&&&&&&&\\
\dfrac{H}{4c} & \dfrac{dH}{4c^2} & -\dfrac{H^2}{4c^2} & -\dfrac{H}{4c} &
\dfrac{dH}{4c^2} & -\dfrac{H^2}{4c^2} & 0 & 0 & 4-\dfrac{d^2}{c^2}\\
\end{array}\right)
\end{equation}

\subsection{Bent Stripes}

Here we compute the Hamiltonian of the ground state consisting of
maximally-zigzagging stripes. It has a unit cell of two particles, $0$
and $1$, and we set the position of particle $0$ as the origin and 
the $x$-direction to run along the line connecting particles $0$
and $1$. Particle $0$ represents the particles with odd $m$, and
particle $1$ represents the particles with even $m$, hence we set
for particle $0$, $m=2t-1$, and for particle $1$, $m=2t$, where $1\leq
t\leq L/2$. 
The positions of the two particles in the unit cell and theirs $8$
neighbors are listed in Table S2. 
\begin{table}[h!]
\label{tab:bent}
\begin{center}
\begin{tabular}{|c|*{4}{c|}}
\hline
\mbox{\it l} & index & x & y & z\\
\hline
0  & (m,n,p)       & 0  & 0 & $H$\\

1  & (m+1,n,p)     & 2c & 0 & $H$\\

2  & (m+2,n,p-1)   & $2c+c\dfrac{3d^2-c^2}{c^2+d^2}$ &
$-d\dfrac{H^2}{c^2+d^2}$ & 0\\

3  & (m+2,n+1,p-1) & $2c+c$ & $d$ & 0\\

4  & (m+1,n+1,p-1) & $c$ & $d$ & 0\\

5  & (m-1,n+1,p)   & $2c\dfrac{c^2-d^2}{c^2+d^2}$ &
$\dfrac{4c^2d}{c^2+d^2}$ & $H$\\

6  & (m-1,n,p-1)   & $-c\dfrac{3d^2-c^2}{c^2+d^2}$ &
$d\dfrac{H^2}{c^2+d^2}$ & 0\\

7  & (m-1,n-1,p-1) & $-c$ & $-d$ & 0\\

8  & (m+1,n-1,p-1) & $c$  & $-d$ & 0\\

9  & (m+2,n-1,p)   & $2c-2c\dfrac{c^2-d^2}{c^2+d^2}$ &
$-\dfrac{4c^2d}{c^2+d^2}$ & $H$\\
\hline
\end{tabular}
\hspace{0.5cm}
\begin{tabular}{|c|*{3}{c|}}
\hline
particles & dx & dy & dz\\
\hline
1,0  & 2c & 0 & 0\\

4,0  & $c$ & $d$ & -H\\

5,0  & $2c\dfrac{c^2-d^2}{c^2+d^2}$ &
$\dfrac{4c^2d}{c^2+d^2}$ & 0\\

6,0  & $-c\dfrac{3d^2-c^2}{c^2+d^2}$ &
$d\dfrac{H^2}{c^2+d^2}$ & -H\\

7,0  & $-c$ & $-d$ & -H\\

8,0  & $c$  & $-d$ & -H\\

2,1  & $c\dfrac{3d^2-c^2}{c^2+d^2}$ &
$-d\dfrac{H^2}{c^2+d^2}$ & -H\\

3,1  & $c$ & $d$ & -H\\

4,1  & $-c$ & $d$ & -H\\

8,1  & $-c$  & $-d$ & -H\\

9,1  & $-2c\dfrac{c^2-d^2}{c^2+d^2}$ &
$-\dfrac{4c^2d}{c^2+d^2}$ & 0\\
\hline
\end{tabular}
\caption{Indices and positions of the particles $0$ and $1$ and their
  eight neighbours $l=2,...,9$ (left) and distances between the
  neighbouring particles and $0$ and $1$ particles (right) in the unit
  cell of the maximally zigzagging stripe configuration.} 
\end{center}
\end{table}

The Hamiltonian is
\begin{equation}\label{equ:HamiltonB}
\begin{array}{ll}
\mathcal{H}= & \dfrac{U_0}{4\delta^2}\sum_{t,n,p}\Big\{-4c(dr_{10}+dr_{40}+dr_{50}+dr_{60}+dr_{70}+dr_{80}+dr_{21}+dr_{31}+dr_{41}+dr_{81}+dr_{91})\\
&\\
&+(dr_{10}^2+dr_{40}^2+dr_{50}^2+dr_{60}^2+dr_{70}^2+dr_{80}^2+dr_{21}^2+dr_{31}^2+dr_{41}^2+dr_{81}^2+dr_{91}^2)\Big\}+\mathcal{H}_{wall}
\end{array}
\end{equation}
Absorbing the terms with indices $60,70,80,01,81,91$ in the terms with
indices $10,40,50,21,31,41$, the Hamiltonian becomes
\begin{equation}
\begin{array}{ll}
\mathcal{H}= &
\dfrac{U_0}{2\delta^2}\displaystyle\sum_{t,n,p}\Bigg\{\dfrac{3c^4+5d^4}{2(c^2+d^2)^2}(u_0^2+u_1^2)-u_0u_1-\dfrac{1}{4}(u_0u_4+u_1u_3)+\dfrac{4c^2d^2}{(c^2+d^2)^2}u_0u_5\\
&\\
&
+\dfrac{d(3c^4-4c^2d^2+d^4)}{c(c^2+d^2)^2}(u_0v_0+u_1v_1)-\dfrac{d}{4c}(u_0v_4+u_1v_3-u_1v_4+u_3v_1+u_4v_0-u_4v_1)\\
&\\
&
-\dfrac{2cd(c^2-d^2)}{(c^2+d^2)^2}(u_0v_5+u_5v_0)+\dfrac{(-c^4+6c^2d^2-9d^4)}{4(c^2+d^2)^2}u_1u_2+\dfrac{d(3d^2-c^2)(3c^2-d^2)}{4c(c^2+d^2)^2}\\
&\\
&
\cdot(u_1v_2+u_2v_1)+\dfrac{H}{4c}\sign(dz_0-H/2)(u_0w_4+u_1w_3-u_1w_4+u_3w_1+u_4w_0-u_4w_1)\\
&\\
& +\dfrac{H(3d^2-c^2)}{4c(c^2+d^2)}\sign(dz_0-H/2)(u_1w_2+u_2w_1)+\dfrac{d^2(7c^4+d^4)}{2c^2(c^2+d^2)^2}(v_0^2+v_1^2)\\
&\\
&
-\dfrac{d^2}{4c^2}(v_0v_4+v_1v_3+v_1v_4)-\dfrac{4c^2d^2}{(c^2+d^2)^2}v_0v_5+\dfrac{d^2(-9c^4+6c^2d^2-d^4)}{4c^2(c^2+d^2)^2}v_1v_2\\
&\\
&
+\dfrac{Hd}{4c^2}\sign(dz_0-H/2)(v_0w_4+v_1w_3+v_1w_4+v_3w_1+v_4w_0+v_4w_1)\\
&\\
&-\dfrac{Hd(3c^2-d^2)}{4c^2(c^2+d^2)}\sign(dz_0-H/2)(v_1w_2+v_2w_1)+(2-\dfrac{d^2}{2c^2})(w_0^2+w_1^2)\\
&\\
&-\dfrac{H^2}{4c^2}(w_0w_4+w_1w_2+w_1w_3+w_1w_4)\Bigg\}
\end{array}
\end{equation}
In the following we specify the expression of the matrix $A$ for 1,2
and 3 particles free to move, $A_b^{0}$,$A_b^{01}$ and $A_b^{014}$
respectively, where the subscript $b$ stays for bent and the
superscript indicates the particles as specified in Table S2, which are
free to move.

\begin{equation}
A_b^0=\left(\begin{array}{ccc}
A_{11} & A_{12} & A_{13}\\
&&\\
A_{12} & A_{22} & A_{23}\\
&&\\
A_{13} & A_{23} & A_{33}\\
\end{array}\right)
\end{equation}
\begin{equation}
A_b^{01}=\left(\begin{array}{cccccc}
A_{11} & A_{12} & A_{13} & -1 & 0 & 0\\
&&&&&\\
A_{12} & A_{22} & A_{23} & 0 & 0 & 0\\
&&&&&\\
A_{13} & A_{23} & A_{33} & 0 & 0 & 0\\
&&&&&\\
-1 & 0 & 0 & A_{11} & A_{12} & -A_{13}\\
&&&&&\\
0 & 0 & 0 & A_{12} & A_{22} & -A_{23}\\
&&&&&\\
0 & 0 & 0 & -A_{13} & -A_{23} & A_{33}\\
\end{array}\right)
\end{equation}
\begin{equation}
A_b^{012}=\left(\begin{array}{ccccccccc}
A_{11} & A_{12} & A_{13} & -1 & 0 & 0 & -\dfrac{1}{4} & A_{18} & A_{19}\\
&&&&&&&&\\
A_{12} & A_{22} & A_{23} & 0 & 0 & 0 & A_{18} & A_{28} & A_{29}\\
&&&&&&&&\\
A_{13} & A_{23} & A_{33} & 0 & 0 & 0 & A_{19} & A_{29} & A_{39}\\
&&&&&&&&\\
-1 & 0 & 0 & A_{11} & A_{12} & -A_{13} & -\dfrac{1}{4} & -A_{18} & -A_{19}\\
&&&&&&&&\\
0 & 0 & 0 & A_{12} & A_{22} & -A_{23} & -A_{18} & A_{29} & A_{29}\\
&&&&&&&&\\
0 & 0 & 0 & -A_{13} & -A_{23} & A_{33} & -A_{19} & A_{29} & A_{39}\\
&&&&&&&&\\
-\dfrac{1}{4} & A_{18} & A_{19} & -\dfrac{1}{4} & -A_{18} & -A_{19} & A_{11} & A_{12} & -A_{13}\\
&&&&&&&&\\
A_{18} & A_{28} & A_{29} & -A_{18} & A_{28} & A_{29} & A_{12} & A_{22} & -A_{23}\\
&&&&&&&&\\
A_{19} & A_{29} & A_{39} & -A_{19} & A_{29} & A_{39} & -A_{13} & -A_{23} & A_{33}\\
\end{array}\right)
\end{equation}
The elements of matrices $A_b^0$,$A_b^{01}$ and $A_b^{014}$ are
specified in the following
\noindent\begin{equation}
\begin{array}{ll}\left\{
\begin{array}{lll}
A_{11} & = & \dfrac{3c^4+5d^4}{(c^2+d^2)^2}\\
&&\\
A_{12} & = & \dfrac{d(c^2-d^2)H^2}{c(c^2+d^2)^2}\\
&&\\
A_{13} & = & \dfrac{(d^2-c^2)H}{2c(c^2+d^2)}\\
&&\\
A_{18} & = & -\dfrac{d}{4c}\\
&&\\
A_{19} & = & \dfrac{H}{4c}\\ 
\end{array}\right.
& \ \ \ \ \ \ \ \ \ \
\left\{\begin{array}{lll}
A_{22} & = & \dfrac{d^2(7c^4+d^4)}{c^2(c^2+d^2)^2}\\
&&\\
A_{23} & = & \dfrac{d(d^2-c^2)H}{2c^2(c^2+d^2)}\\
&&\\
A_{33} & = & 4-\dfrac{d^2}{c^2}\\
&&\\
A_{28} & = & -\dfrac{d^2}{4c^2}\\
&&\\
A_{29} & = & \dfrac{dH}{4c^2}\\
&&\\
A_{39} & = & -\dfrac{H^2}{4c^2}\\
\end{array}\right.
\end{array}
\end{equation}
Using the expression for the entropy in Eq.(\ref{equ:entropy}), we can
obtain the entropy difference per particle between straight and bent
configurations: $\Delta s=1/n\cdot\log(\|A_b\|/\|A_s\|)$, as shown in
Fig.~\ref{fig:entropy_U}a for 1, 2 and 3 particles free to move.

\vspace{0.3cm}
\begin{center}
{\bf SIII. Soft spheres interacting through the square-shoulder
  repulsive potential}
\end{center}

One of the simplest soft potentials close to the hard-sphere model
is the square-shoulder repulsive potential which describes particles
with a hard core surrounded by a soft corona.
This model can have a very rich behavior and it has been shown to
develop pattern formation \cite{malescio2003}, different mesophases
\cite{glaser2007} and quasi-crystals \cite{dotera2014}.    
In the case of one sphere free to move, the entropy of such soft
spheres system in the canonical ensemble can be evaluated through the
partition function assuming the following soft inter-particle
potential
\begin{equation}
\phi_{soft}(r)=\left\{\begin{array}{cl}
\infty & \mbox{for $r\le a_H$} \\
\phi_0 & \mbox{for $a_H<r\le a_S$} \\
0 & \mbox{for $r>a_S$}
\end{array}\right.
\end{equation}
where $a_H$ and $a_S$ are the hard and soft radii respectively, 
and $\phi_0$ is the soft potential strength.  
In computing the partition function one can split the integral on the
coordinate {\it r} of the center of the free sphere to the following
three regions: 
$|r-r_i|\le a_H$, $a_H<|r-r_i|<a_S$ and $|r-r_i|\ge a_S$ for all $r_i$
where $r_i$ are the coordinates of the six spheres and the wall which
confine the free sphere. 
The integral on the first region does not contribute to the partition
function while the integrals on the second and third regions contribute
the terms $e^{-\beta\phi_0}\Delta V_{1,\rho_H,\rho_S}$ and
$V_{1,\rho_S}$ respectively, where $\rho_{H,S}=(4/3)\pi a_{H,S}^3
N/V_0$ and $\Delta V_{1,\rho_H,\rho_S}=V_{1,\rho_H}-V_{1,\rho_S}$.  
Using the expression of the entropy in the left-hand side of 
Eq.~(\ref{equ:entropy}) we obtain
\begin{equation}
S_{1,\rho_H,\rho_S}^{soft}=\dfrac{\beta\phi_0
  e^{-\beta\phi_0}\Delta
  V_{1,\rho_H,\rho_S}}{V_{1,\rho_S}+e^{-\beta\phi_0}\Delta
  V_{1,\rho_H,\rho_S}}+\ln(V_{1,\rho_S}+e^{-\beta\phi_0}\Delta V_{1,\rho_H,\rho_S}).
\end{equation}
The thermal factor, which for soft spheres couples with space
variables, is $0<e^{-\beta\phi_0}<1$.
Because $V_{1,\rho_S}^{bent}>V_{1,\rho_S}^{straight}$ and $\Delta
V_{1,\rho_H,\rho_S}^{bent}>\Delta V_{1,\rho_H,\rho_S}^{straight}$, we have
that $S_{1,\rho_H,\rho_S}^{soft,\ \ bent}>S_{1,\rho_H,\rho_S}^{soft,\ \ straight}$
so that bent stripes are still favored with respect to straight
stripes in the case of one sphere free to move interacting through a
soft-shoulder potential.  
In the general case of $n$ free spheres, the computation of
$S_{n,\rho_H,\rho_S}^{soft}$ is complicated due to the presence of series of
powers of $\phi_0$, but, as for the hard-spheres model, we expect that
increasing $n$ will not change the preference for bent stripes with
respect to straight stripes.

\vspace{0.3cm}
\begin{center}
{\bf SIV. Asymmetric potential}
\end{center}

\begin{figure}[h!]
\includegraphics[clip=true,width=11cm,height=5.82cm]{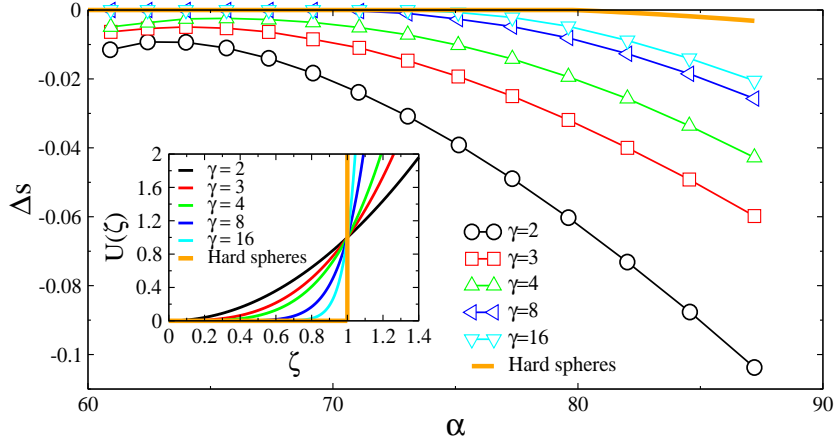}  
\caption{Entropy difference per particle $\Delta s = \Delta S/n$
  between straight and bent stripes configurations for $n=1$ for the
  generalized repulsive harmonic potential
  $U_r^{\gamma}(\zeta)$. Inset: $U_r^{\gamma}(\zeta)$.}    
\label{fig:Repulsive}  
\end{figure}

As an asymmetric potential we first consider the generalized repulsive
potential $U_r^{\gamma}(\zeta)=U_r^0\zeta^\gamma\cdot\theta(\zeta)$
which reduces to the repulsive harmonic potential for $\gamma=2$ and 
gives hard spheres of diameter $2(c-\delta)$ for
$\gamma\rightarrow\infty$. 
$U_r^{\gamma}$ differs by a numerical prefactor from the commonly-used
tunable soft repulsive potential \cite{liu2010,hecke2010,morse2014},
with $dr_0=\sigma_{ij}$ and $dr=r_{ij}$, and the hard-sphere limit
obtained for $\alpha\rightarrow 0$ \cite{liu2010}.
Figure \ref{fig:Repulsive} shows how as $\gamma$ increases, $\Delta
s$ for one particle free to move slowly approaches the hard-sphere
result and always exhibits a preference for bent stripes. 

Now we consider the more general asymmetric power-law potential:
$U_{asy}(\zeta)=U_r^0\zeta^{\gamma_r}\theta(\zeta)+U_a^0(-\zeta)^{\gamma_a}\theta(-\zeta)$
where $\zeta=(dr_0-dr)/(2\delta)$, with $U_r^0$, $U_a^0$,
$\gamma_r$, $\gamma_a$ constants. The subscripts $r$ and $a$ denote
repulsive and attractive respectively. 
In principle, at finite $T$, we should compare the free energy of the
competing configurations to establish the thermodynamically stable
phase. 
If the Hamiltonian of the system includes only quadratic terms in the
canonical variables, as for the deformable antiferromagnetic Ising
model or particles interacting through the harmonic potential, the
internal energy of straight and bent configurations is the same, as
follows from the equipartition theorem, and the difference in the free
energy comes only from entropy. 
For finite $T$, the internal energy with the asymmetric
potential $U_{asy}$, for which we can not apply the equipartition
theorem, could grow differently with temperature for fluctuations
around different ground-state configurations. 
In the following we will show that to establish which is the more
stable configuration for particles interacting through $U_{asy}$, we
can compare just the entropy because for small $T$ the internal energy
of the different configurations is the same, as for the harmonic
potential.  

For a harmonic system the equipartition theorem leads to the result
that each degree of freedom contributes $\frac{1}{2}kT$ to the
internal energy. 
For anharmonic potentials, as $U_{asy}$, it is not obvious that the
energy grows with $T$ exactly in the same functional form for all
competing ground-state configurations. 
To check this for $U_{asy}$, we start by considering a simple example
of a one-dimensional asymmetric power-law potential of the form  
\begin{eqnarray}
U(x) = \left\{ \begin{array}{ll} U_a^0 (-x)^{\gamma_a} & x<0 \\  U_r^0
  x^{\gamma_r} & x>0 \end{array}\right. . 
\end{eqnarray}

Its canonical partition function is given by
\begin{eqnarray}
Z = \int_{-\infty}^{\infty}\exp\left[-\beta U(x)\right]dx =
\int_{-\infty}^0 \exp\left[-\beta U_a^0 (-x)^{\gamma_a}\right]dx +
\int_0^{\infty} \exp\left[-\beta U_r^0 x^{\gamma_r}\right]dx . 
\end{eqnarray}
By change of variables we see that this may be written as
\begin{eqnarray}
Z = C(\gamma_a) \left(\beta U_a^0\right)^{-1/\gamma_a} +
C(\gamma_r) \left(\beta U_r^0\right)^{-1/\gamma_r} , 
\end{eqnarray}
where the prefactors $C(\gamma)$ will be soon shown to be irrelevant.  

The internal energy is given by
\begin{eqnarray}
\langle U\rangle = - \frac{1}{Z} \frac{\partial Z}{\partial \beta} =
\frac{\dfrac{C(\gamma_a)}{\gamma_a \beta} \left(\beta U_a^0
  \right)^{-1/\gamma_a} +  \dfrac{C(\gamma_r)}{\gamma_r \beta}
  \left(\beta U_r^0 \right)^{-1/\gamma_r} }{
  C(\gamma_a)\left(\beta U_a^0 \right)^{-1/\gamma_a} +
  C(\gamma_r)\left(\beta U_r^0 \right)^{-1/\gamma_r} } . 
\end{eqnarray} 
where $\langle\rangle$ refers to the canonical ensemble average.
We use the convention that Bolztmann's constant is set to unity, thus
substitute $\beta=1/T$ and write this as 
\begin{eqnarray}
\langle U\rangle = \frac{\dfrac{C(\gamma_a)}{\gamma_a}
  (U_a^0)^{-1/\gamma_a} T^{1+1/\gamma_a} +
  \dfrac{C(\gamma_r)}{\gamma_r} (U_r^0)^{-1/\gamma_r} 
  T^{1+1/\gamma_r} }{ C(\gamma_a) (U_a^0)^{-1/\gamma_a} T^{1/\gamma_a} +
  C(\gamma_r) (U_r^0)^{-1/\gamma_r} T^{1/\gamma_r} } . 
\end{eqnarray}
Since $\gamma_a , \gamma_r > 1$, as $T \rightarrow 0$, $T^{1/\gamma_a}
, T^{1/\gamma_r} \rightarrow 0$. We are interested in the case that
$\gamma_a \neq \gamma_r$. We will denote the smaller and larger of
these two exponents as $\gamma_1 = max(\gamma_a,\gamma_r) , \gamma_2 =
min(\gamma_a,\gamma_r)$. That is, $\gamma_1 > \gamma_2$. Now, as $T
\rightarrow 0$, $T^{1/\gamma_1} \gg T^{1/\gamma_2}$, and thus  
\begin{eqnarray}
\langle U\rangle \approx \frac{T}{\gamma_1} .
\end{eqnarray}
Namely, in the low-temperature limit, the internal energy grows with
increasing temperature in a manner that depends only on the exponent
of the stiffer side of the interaction (attractive or repulsive), and
does not depend on the prefactors $U_a^0,U_r^0$. 
From the other hand, if we apply the generalised equipartition
function to $U(x)$, that is $\langle x\partial U(x)/\partial
x\rangle=T$, we obtain the same result for $\langle U\rangle$.
If we apply the generalised equipartition function to
$U_{asy}(\zeta)$, we can obtain the same relation as for $\langle
U\rangle$, but with a different prefactor.  
In other words, we have that for $U_{asy}(\zeta)$ the only difference
in free energy between straight and bent stripes configurations comes
from the entropy difference.   

If we fix $U_r^0=U_a^0=1$, we find, computing by numerical
integration the canonical partition function, that there is a value
of $\gamma_a>\gamma_r$ ($\gamma_a=\gamma_r\cdot\kappa(\alpha)$, with
$0<1/\kappa<1$) for which $\Delta s<0$. See Fig.~\ref{fig:entropy_U}.

If we consider the case of $\gamma_r=\gamma_a$ and change only the
prefactors $U_r^0$ and $U_a^0$, we do not find a stripe inversion
(that is we never find $\Delta s<0$ for any $U_r^0$ and $U_a^0$).  
This can also be demonstrated by showing that the attractive
component of the potential obtained by changing only the prefactor
$U_a^0$ is always bigger than the attractive componend obtained by
changing only the exponent $\gamma_a$ with respect to $\gamma_r$ (even
for $\gamma_a/\gamma_r<\kappa$ for which $\Delta s>0$).     
Without loss of generality we fix $U_r^0=1$ and change $U_a^0$.  
In the calculation of the partition function $Z$, we have to integrate
over a product of exponentials of the form $\exp(-\beta
U_r^0\zeta^{\gamma_r})$ and $\exp[-\beta U_a^0(-\zeta)^{\gamma_a}]$.
Because we consider $T\rightarrow 0$, that is $\beta\rightarrow
\infty$, the main contribution to $Z$ comes from small values of
$\zeta$. 
It results that, once we fix the value of $U_a^0$, for
$\gamma_a/\gamma_r>1$ there is a value of
$\zeta=\zeta_0=\zeta_0(\gamma_a/\gamma_r)>0$ such that for
$\zeta<\zeta_0$ we have $\zeta_0^{\gamma_a}<U_a^0\zeta_0^{\gamma_r}$,
that is $\zeta_0^{\gamma_a/\gamma_r}<U_a^0$, for any fixed $U_a^0$ and 
$\gamma_a/\gamma_r>1$ (in particular also for
$1<\gamma_a/\gamma_r<\kappa$, that is for a value of
$\gamma_a/\gamma_r$ that is still too small to cause the inversion of
$\Delta s$).

\end{document}